\documentclass[10pt,twocolumn]{article}

\usepackage[margin=0.75in]{geometry}
\usepackage[T1]{fontenc}
\usepackage[utf8]{inputenc}
\usepackage{newtxtext,newtxmath}
\usepackage{amsmath}
\usepackage{tikz-cd}
\usepackage{graphicx}
\usepackage{bm}
\usepackage{xcolor}
\usepackage[colorlinks=true,citecolor=blue,linkcolor=blue,urlcolor=blue,hypertexnames=false]{hyperref}
\usepackage{abstract}
\usepackage{titlesec}
\usepackage{authblk}
\usepackage{setspace}
\usepackage{placeins}
\usepackage{cuted}
\usepackage{capt-of}
\usepackage{float}
\usepackage[backend=biber,style=chem-acs,sorting=none,doi=true,url=false]{biblatex}
\usepackage{siunitx}
\DeclareSIUnit{\counts}{counts}
\DeclareSIUnit{\angstromunit}{\text{\AA}}
\usepackage[version=4]{mhchem}
\usepackage{makecell}
\renewcommand{\thesection}{\Roman{section}}
\titleformat{\section}{\normalfont}{\thesection.}{0.5em}{\MakeUppercase}
\titleformat{\subsection}{\normalfont\bfseries}{\thesubsection.}{0.5em}{}
\titlespacing*{\section}{0pt}{*2}{*1}
\titlespacing*{\subsection}{0pt}{*1.5}{*0.75}
\makeatletter
\renewcommand{\@maketitle}{%
  \newpage
  \null
  \vspace*{2em}%
  \begin{center}%
    \begin{minipage}{\textwidth}%
      \begin{center}%
        {\normalfont\bfseries\LARGE \@title\par}%
        \vspace*{1.5em}%
        {\@author\par}%
        \vspace*{1.5em}%
      \end{center}%
    \end{minipage}%
  \end{center}%
}
\makeatother

\makeatletter
\def\@font@warning#1{}
\makeatother
\title{Color-Center-Compatible Freestanding Diamond Directional Couplers for Quantum Photonics}

\author[1,*]{Colin Sauerzapf}
\author[1]{Tom Jäger}
\author[1]{Jonathan Enßlin}
\author[1]{Oliver von Berg}
\author[1,2]{Vladislav Bushmakin}
\author[1]{Roman Kolesov}
\author[1]{Rainer Stöhr}
\author[1,$\dagger$]{Vadim Vorobyov}
\author[1,2]{Jörg Wrachtrup}

\affil[1]{3rd Institute of Physics, University of Stuttgart, Stuttgart 70569, Germany}
\affil[2]{Max Planck Institute for Solid State Research, Stuttgart 70569, Germany}
\affil[*]{E-mail: \href{mailto:colin.sauerzapf@pi3.uni-stuttgart.de}{colin.sauerzapf@pi3.uni-stuttgart.de}}
\affil[$\dagger$]{E-mail: \href{mailto:v.vorobyov@pi3.uni-stuttgart.de}{vadim.vorobyov@pi3.uni-stuttgart.de}}

\begin{document}
\begin{refsection}
  \twocolumn[
    \begin{@twocolumnfalse}
      \maketitle
      \begin{abstract}
        \noindent Freestanding all-diamond color-center photonics is a promising platform for optical integration of spin-based quantum defects. Within this geometry, we realize a key building block for quantum-network interconnects: a directional coupler that acts as an on-chip beam splitter.
We design and simulate directional couplers with triangular cross sections using eigenmode and finite-difference time-domain simulations and target near-50:50 splitting at visible wavelengths. We fabricate the devices directly from bulk single-crystal diamond by angled oxygen reactive-ion-beam etching followed by a dry post-release hard-mask removal process. Room-temperature measurements at \mbox{$\lambda_0\approx\SI{637}{\nano\meter}$} yield a mean coupling ratio of \mbox{$C^\mathrm{meas}=\SI{46\pm16}{\percent}$}. Finally, we integrate SnV$^{-}$ centers into the nanophotonic structures and observe near-lifetime-limited optical linewidths and coherent optical Rabi oscillations without post-fabrication annealing, identifying the platform as a viable route towards integrated diamond quantum photonics.
      \end{abstract}
      {\small\noindent\textbf{Keywords:} \textit{diamond, directional couplers, color centers, tin-vacancy centers, integrated quantum photonics}\par}
      \vspace{1.2em}
    \end{@twocolumnfalse}
  ]

  \section{Introduction}

\noindent Diamond has emerged as a leading solid-state platform for quantum photonics because it combines a wide electronic bandgap of \SI{5.5}{\electronvolt} with a chemically robust host lattice for optically addressable point defects \cite{awschalom_quantum_2018,aharonovich_diamond_2011}.
Color centers, particularly NV centers, are attractive photon-interfaced quantum memories and have matured as candidates for quantum-network and quantum-computing applications \cite{ruf_quantum_2021, pompili_realization_2021, childress_diamond_2013}.
Beyond NV centers, group-IV vacancy centers such as SiV$^{-}$, GeV$^{-}$, and especially SnV$^{-}$ offer narrow optical transitions combined with inversion-symmetric defect structures \cite{iwasaki_tin-vacancy_2017, thiering_ab_2018}. This symmetry suppresses sensitivity to electric-field noise and makes these emitters particularly robust against charge fluctuations. As a result, they can be integrated into nanophotonic structures \cite{rugar_quantum_2021, rugar_narrow-linewidth_2020} and may enable efficient remote quantum nodes \cite{knaut_entanglement_2024}.

Scalable diamond-based quantum architectures, however, require a complete quantum-photonic toolbox, ranging from quantum emitters to passive and active photonic components that are ideally integrated on a single chip \cite{giordani_integrated_2023,kennard_-chip_2013, pezzagna_quantum_2021}.
Among these components, a 50:50 beam splitter is central because it is required in many remote-entanglement distribution schemes \cite{ruf_quantum_2021,childress_diamond_2013}.
Implementing such components on chip would enable photonic interconnects between color-center-based qubit registers, providing a route towards modular scaling of solid-state quantum processors \cite{kennard_-chip_2013,childress_diamond_2013}.

Realizing scalable integrated diamond-photonic circuits, however, remains technologically challenging.

\begin{figure}[H]
  \centering
  \includegraphics[width=\columnwidth]{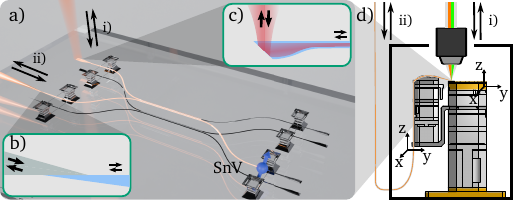}
\caption{\textbf{a)} Schematic of a freestanding all-diamond directional coupler with an embedded SnV$^{-}$ center, illustrating the intended routing of waveguide-coupled emission between two output arms by evanescent coupling. One output is collected out of plane via a total internal reflection (TIR) coupler (i), while the other is coupled to a tapered optical fiber (ii). \textbf{b)} Illustration of tapered-fiber coupling, as discussed in Ref.~\cite{burek_fiber-coupled_2017}. \textbf{c)} Illustration of the TIR coupler, in which guided light is redirected out of plane by total internal reflection at the undercut diamond interface and collected with an objective. \textbf{d)} Schematic of the low-temperature measurement setup used to characterize emitters in the photonic chip at \mbox{$T=\SI{4}{\kelvin}$}. The cryostat provides a base temperature of \mbox{$T\gtrsim\SI{70}{\milli\kelvin}$}. Optical access to the chip is provided both confocally from above (i) and via a nanopositioner-mounted tapered fiber (ii).}
\label{fig:introduction-platform}
\end{figure}

\noindent Hybrid approaches, in which diamond emitters are coupled to photonic structures made from other materials, can benefit from mature fabrication platforms but introduce additional interfaces, alignment constraints, and integration complexity \cite{katsumi_recent_2025, riedel_efficient_2023}. In addition, differences in thermal expansion and heat dissipation can make hybrid structures susceptible to strain during cryogenic packaging and operation. Thin-film diamond photonics offers a more direct route to integrated devices \cite{gao_directional_2018,ding_high-q_2024}, yet scalable access to high-quality single-crystal diamond membranes remains limited \cite{guo_tunable_2021}.

An alternative strategy is to fabricate freestanding photonic structures directly from bulk single-crystal diamond \cite{burek_free-standing_2012,burek_high_2014,atikian_freestanding_2017}.
This approach preserves the high-quality diamond host while eliminating the need for membrane transfer and maintaining thermal contact with the bulk diamond. However, it requires a fabrication process capable of reliably releasing mechanically delicate nanostructures while preserving their optical geometry. Both quasi-isotropic and angled etching have been explored to realize such freestanding structures from bulk diamond \cite{atikian_freestanding_2017,chia_development_2022, khanaliloo_high-_2015}.
In previous work, inverse design was used to realize beam splitters based on quasi-isotropically etched diamond structures \cite{dory_inverse-designed_2019}.
Applying the same approach to angled-etched devices is less straightforward because the triangular waveguide cross section makes the problem intrinsically three-dimensional and therefore not easily reducible to a two-dimensional design.
At the same time, angled etching is largely independent of the crystal axis, providing additional freedom in the layout of freestanding photonic structures.

In this work, we realize freestanding diamond directional couplers, as depicted in Fig.~\ref{fig:introduction-platform}a, fabricated directly from bulk single-crystal diamond by electron-beam lithography, oxygen reactive-ion-beam etching (RIBE) \cite{atikian_freestanding_2017}, and dry post-release hard-mask removal. To optically interface with the chip, we combine established tapered-fiber coupling (Fig.~\ref{fig:introduction-platform}a,b) \cite{burek_fiber-coupled_2017} with a total internal reflection (TIR) coupler that redirects guided light out of plane, as illustrated in Fig.~\ref{fig:introduction-platform}a,c.

Through eigenmode and finite-difference time-domain (FDTD) simulations, fabrication, and optical characterization, we demonstrate near-balanced directional coupling in freestanding diamond waveguides at visible wavelengths relevant to SnV$^{-}$ centers. We further characterize SnV$^{-}$ centers embedded in these freestanding nanophotonic structures using the low-temperature setup depicted in Fig.~\ref{fig:introduction-platform}d. Together, these results establish dry-processed, bulk-derived freestanding diamond photonics as a promising route towards integrated quantum-photonic circuits with group-IV color centers.

  \section{Fabrication and design}
\subsection*{Fabrication}
\begin{figure}[H]
  \centering
  \includegraphics[width=\columnwidth]{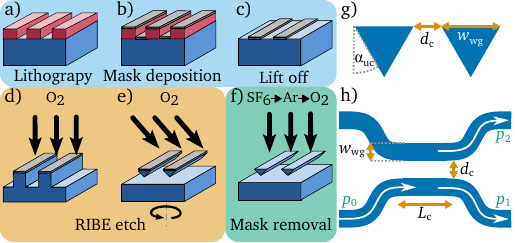}
\caption{Fabrication process from bulk diamond to a freestanding structure. \textbf{a)} The desired layout is patterned into a positive resist (red) by electron-beam lithography. \\ \textbf{b)} Chromium and titanium are then deposited to form the hard-mask material. \textbf{c)} The resist is subsequently removed by lift-off. The mask is then transferred into the bulk diamond by highly directional \ce{O2} RIBE. \textbf{d)} The first etch is surface-normal. \textbf{e)} The second is performed with an ion beam incident at \SI{45}{\degree} while the sample continuously rotates in plane. \textbf{f)} Finally, the hard mask is removed by consecutive dry \ce{SF6}, \ce{Ar}, and \ce{O2} ICP-RIE steps, revealing the freestanding diamond-only structures. Detailed fabrication parameters are provided in Supplementary Information \cite{supplementary-information}, Sec.~S1. \textbf{g)} Cross-sectional schematic of the coupling region indicating the undercut angle $\alpha_\mathrm{uc}$, waveguide separation $d_\mathrm{c}$, and waveguide width $w_\mathrm{wg}$. \textbf{h)} Top-view schematic of the directional coupler, additionally showing the input and output ports $p_0$, $p_1$, and $p_2$, as well as the coupling length $L_\mathrm{C}$.}
\label{fig:fabrication-steps}
\end{figure}

A mask is patterned into a layer of spin-coated positive resist on single-crystal diamond by electron-beam lithography (EBL), as shown in Fig.~\ref{fig:fabrication-steps}a. After development, a hard mask consisting of \SI{1}{\nano\meter} chromium (\ce{Cr}) and \SI{80}{\nano\meter} titanium (\ce{Ti}) is deposited (Fig.~\ref{fig:fabrication-steps}b), followed by lift-off using a resist remover (Fig.~\ref{fig:fabrication-steps}c). The mask pattern is transferred into the bulk diamond in two RIBE steps: a surface-normal oxygen (\ce{O2}) plasma etch, perpendicular to the diamond surface (Fig.~\ref{fig:fabrication-steps}d), followed by an angled \ce{O2} etch in which the sample is tilted to \mbox{$\alpha_\mathrm{e}=\SI{45}{\degree}$} and rotated in plane (Fig.~\ref{fig:fabrication-steps}e). This angled etch undercuts the structures from all in-plane directions with respect to the surface normal, releasing arbitrarily oriented freestanding structures from the bulk diamond.

After release, the hard mask is removed using a dry inductively coupled plasma reactive-ion-etching (ICP-RIE) sequence, as shown in Fig.~\ref{fig:fabrication-steps}f. This dry post-release hard-mask removal is enabled by the high selectivity of \ce{SF6} plasma for titanium over diamond, allowing the titanium mask layer to be etched while leaving the diamond structure largely unaffected. The remaining chromium layer is subsequently removed by an \ce{Ar} plasma step, followed by a soft \ce{O2} ICP etch with no radio-frequency bias applied to the sample electrode, to further polish and oxygen-terminate the diamond surface. Avoiding wet processing after release mitigates meniscus-force-induced damage and thereby enables more delicate freestanding structures. Detailed process parameters are provided in Supplementary Information \cite{supplementary-information}, Sec.~S1.

\begin{figure*}[t]
\centering
\includegraphics[width=\textwidth]{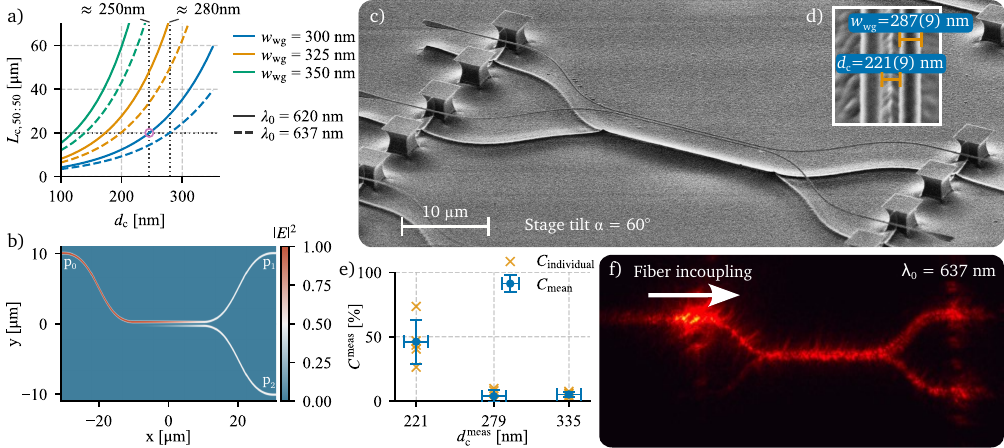}
\caption{\textbf{a)} Coupling length $L_\mathrm{C,50:50}$ for 50:50 splitting as a function of waveguide separation $d_\mathrm{c}$ for different waveguide widths $w_\mathrm{wg}$ and wavelengths $\lambda_0$, extracted from eigenmode simulations. \textbf{b)} FDTD simulation of a directional coupler with coupling length \mbox{$L_\mathrm{C,50:50}=\SI{20}{\micro\meter}$}, waveguide separation \mbox{$d_\mathrm{c}=\SI{250}{\nano\meter}$}, and waveguide width \mbox{$w_\mathrm{wg}=\SI{300}{\nano\meter}$}. \textbf{c)} Scanning electron microscope (SEM) image of a fabricated directional coupler. \textbf{d)} Extraction of the fabricated waveguide width and waveguide separation from an SEM image. The target design parameters were \mbox{$w_\mathrm{wg}=\SI{300}{\nano\meter}$} and \mbox{$d_\mathrm{c}=\SI{250}{\nano\meter}$}. \textbf{e)} Measured coupling ratio $C^\mathrm{meas}$ as a function of waveguide separation $d_\mathrm{c}^\mathrm{meas}$, obtained from optical transmission measurements of fabricated directional couplers at \mbox{$\lambda_0\approx\SI{637}{\nano\meter}$}. Orange markers show the coupling ratios of individual devices, while blue markers indicate the mean coupling ratio for each measured waveguide separation. \textbf{f)} Top-view camera image of a directional coupler with light coupled into the upper-left input port using a tapered fiber, visualizing the splitting behavior.}
\label{fig:design-fabrication}
\end{figure*}

\subsection*{Photonic design}
To narrow the design space for directional couplers while accounting for fabrication constraints, we first simulate the optical supermodes of the coupling region with an eigenmode solver (Lumerical MODE, Ansys Inc.). From these simulations, we extract the effective refractive indices of the symmetric and antisymmetric TE supermodes and use guided-wave coupled-mode theory to estimate the coupling length \cite{huang_coupled-mode_1994, majety_triangular_2024}. In the idealized, lossless case, light is launched into input port \mbox{$p_0$} with power \mbox{$P_0$} and exits through output ports \mbox{$p_1$} and \mbox{$p_2$} with powers \mbox{$P_1$} and \mbox{$P_2$}, respectively, as depicted schematically in Fig.~\ref{fig:fabrication-steps}h). The coupling length \mbox{$L_\mathrm{C}$} required to obtain a target power fraction \mbox{$C = P_2/P_0 = P_2/\left(P_1+P_2\right)$} is then determined by the effective-refractive-index difference between the two supermodes, \mbox{$\Delta n_\mathrm{eff} = \left|n_\mathrm{eff,s}-n_\mathrm{eff,a}\right|$}:

\begin{equation}
L_\mathrm{C} = \frac{\lambda_0}{\pi \Delta n_\mathrm{eff}} \arcsin\left(\sqrt{C}\right).
\label{eq:coupling}
\end{equation}

Here, \mbox{$\lambda_0$} denotes the free-space wavelength. For a 50:50 power splitting ratio, \mbox{$C=\SI{50}{\percent}$}, this expression simplifies to \mbox{$L_\mathrm{C,50:50} = \lambda_0 / \left(4 \Delta n_\mathrm{eff}\right)$}.

For the directional-coupler simulations, we model the fabricated cross section using an undercut angle, depicted in Fig.~\ref{fig:fabrication-steps}g, of \mbox{$\alpha_\mathrm{uc}^\mathrm{meas}=\SI{42\pm 2}{\degree}$}. This value is estimated from SEM images of reference waveguides fabricated with the process described above and corresponds to an angled etch performed with a sample tilt set value of \mbox{$\alpha_\mathrm{e}=\SI{45}{\degree}$}. With the undercut angle fixed, we sweep two parameters in the coupling region: the waveguide width, \mbox{$w_\mathrm{wg}$}, and the waveguide separation, \mbox{$d_\mathrm{c}$} (see Fig.~\ref{fig:fabrication-steps}h). We perform these simulations at both the wavelength used for the optical transmission measurements, \mbox{$\lambda_\mathrm{0,meas}\approx\SI{637}{\nano\meter}$}, and the SnV$^{-}$ zero-phonon-line (ZPL) wavelength, \mbox{$\lambda_\mathrm{0,SnV}\approx\SI{620}{\nano\meter}$}. The resulting coupling lengths for 50:50 splitting, \mbox{$L_\mathrm{C,50:50}$}, are shown in Fig.~\ref{fig:design-fabrication}a.

The simulations show that narrower waveguides require shorter coupling lengths at fixed separation. This trend occurs because a larger fraction of the optical mode extends into the surrounding vacuum, increasing the evanescent overlap between adjacent waveguides. The final geometry must therefore balance optical performance against fabrication constraints. Smaller separations increase the coupling strength but are more sensitive to fabrication imperfections, whereas longer coupling lengths increase the device footprint, can introduce additional propagation loss and require more mechanical support for the freestanding structure. Based on these trade-offs, we choose a design waveguide width of \mbox{$w_\mathrm{wg}^\mathrm{des}=\SI{300}{\nano\meter}$} and a coupling length of \mbox{$L_\mathrm{C}^\mathrm{des}=\SI{20}{\micro\meter}$}. For this geometry, the eigenmode simulations predict 50:50 splitting for a design separation of \mbox{$d_\mathrm{c,637}^\mathrm{des}=\SI{280}{\nano\meter}$} at \mbox{$\lambda_\mathrm{0,meas}\approx\SI{637}{\nano\meter}$} and \mbox{$d_\mathrm{c,620}^\mathrm{des}=\SI{250}{\nano\meter}$} at \mbox{$\lambda_\mathrm{0,SnV}\approx\SI{620}{\nano\meter}$}. We then convert this eigenmode-optimized geometry into the lithography mask used for fabrication. As a final design check for the SnV$^{-}$-relevant layout, we perform finite-difference time-domain (FDTD) simulations (Lumerical FDTD, Ansys Inc.) at \mbox{$\lambda_\mathrm{0,SnV}\approx\SI{620}{\nano\meter}$} on the corresponding three-dimensional device geometry, obtained by extruding the mask layout. The simulated field distribution is shown in Fig.~\ref{fig:design-fabrication}b.

\subsection*{Fabricated geometry}
Based on the simulated target parameter space, we fabricated directional couplers using the process described above. As shown in the scanning electron microscope (SEM) image in Fig.~\ref{fig:design-fabrication}c, supports outside the coupling region hold the structure in place. In addition, supports connect the freestanding waveguides within the coupling region to stabilize their separation, at the expense of increased optical loss. To account for deviations between the fabricated and nominal design geometries, we fabricate directional couplers with different design waveguide separations, \mbox{$d_\mathrm{c}^\mathrm{des} \in \{\SI{200}{\nano\meter}, \SI{250}{\nano\meter}, \SI{300}{\nano\meter}\}$}. For each designed waveguide separation, the device set comprises five devices with the same nominal waveguide geometry but different numbers of supports within the coupling region, varied from one to five. This systematic support-count variation is used separately from the splitting-ratio measurement to estimate the support-induced attenuation contribution.

SEM inspection reveals deviations from the nominal geometry. For devices with \mbox{$d_\mathrm{c}^\mathrm{des}=\SI{200}{\nano\meter}$}, SEM analysis gives an actual waveguide width of \mbox{$w_\mathrm{wg}^\mathrm{meas}=\SI{287\pm9}{\nano\meter}$} and a waveguide separation of \mbox{$d_\mathrm{c}^\mathrm{meas}=\SI{221\pm9}{\nano\meter}$} (Fig.~\ref{fig:design-fabrication}d). All SEM-based geometry estimates are reported as means of multiple measured distances or angles, with uncertainties given by the corresponding standard deviations. These uncertainties mainly reflect the limited edge definition in the SEM images.

\section{Optical characterization}
The coupling ratio is measured at room temperature in a confocal setup operated at \mbox{$\lambda_0\approx\SI{637}{\nano\meter}$}. Excitation and detection spots can be positioned independently on the sample. Laser light is coupled into the directional-coupler input port \mbox{$p_0$} through a TIR coupler (see Supplementary Information \cite{supplementary-information}, Sec.~S3), and the transmitted powers are collected sequentially from the two output ports. We calculate the measured coupling ratio from the relative output powers as \mbox{$C^\mathrm{meas}=P_2/(P_1+P_2)$}, where \mbox{$P_1$} and \mbox{$P_2$} are the powers detected at output ports \mbox{$p_1$} and \mbox{$p_2$}, respectively. This normalization quantifies only the power splitting between the two outputs; losses are excluded from \mbox{$C^\mathrm{meas}$} and treated separately in Supplementary Information \cite{supplementary-information}, Sec.~S3. From this analysis, we extract a residual device loss of \mbox{$L_\mathrm{residual}=\SI{3.7\pm0.3}{\decibel}$}, an additional support-induced loss in the coupling region of \mbox{$L_\mathrm{support}=\SI{0.6\pm0.1}{\decibel}$} per support, and a combined in- and out-coupling loss through the TIR couplers of \mbox{$L_\mathrm{TIRC}=\SI{12.9\pm0.2}{\decibel}$}.

We fabricated five devices for each designed waveguide separation, all of which were released, optically accessible, and suitable for transmission measurements. As shown in Fig.~\ref{fig:design-fabrication}e, appreciable power transfer to the second output port \mbox{$p_2$} is observed only for devices with a design waveguide separation of \mbox{$d_\mathrm{c}^\mathrm{des}=\SI{200}{\nano\meter}$}. All five devices split light into both output ports. Taking this full device set without further selection, we obtain a mean coupling ratio of \mbox{$C^\mathrm{meas}=\SI{46\pm16}{\percent}$}.

The quoted uncertainty is the standard deviation of the measured coupling ratios across the five fabricated devices. The measured device-to-device spread, as well as the deviation from the simulated design trend, is expected because the coupling ratio is highly sensitive to the fabricated coupling geometry, particularly the waveguide width, waveguide separation, and sidewall angle. The local etch angle on the inner sidewalls of the coupled waveguides may differ from the value extracted from isolated reference waveguides, because mutual shadowing during the angled etch can steepen the sidewalls facing the coupling gap. In addition, SEM-based extraction of the waveguide width and separation has a comparatively large uncertainty due to limited edge contrast and charging effects. Since the evanescent overlap depends strongly on the gap geometry, small variations in separation, width, or sidewall angle can produce substantial changes in \mbox{$C^\mathrm{meas}$}, as discussed in Supplementary Information \cite{supplementary-information}, Sec.~S3. The observed spread is therefore consistent with fabrication-induced geometric variations within the coupling region. Despite this spread, the measurements demonstrate that approximately balanced directional coupling can be achieved in the fabricated freestanding diamond waveguides at a wavelength close to the SnV$^{-}$ zero-phonon line (ZPL) and within the SnV$^{-}$ phonon sideband (PSB).

In a separate room-temperature measurement, we qualitatively visualize the splitting using a configuration similar to that shown schematically in Fig.~\ref{fig:introduction-platform}b: light is injected into the input port of a directional coupler with a tapered fiber, and the scattered light is imaged from above with a camera. The resulting image is shown in Fig.~\ref{fig:design-fabrication}f.

  \section[SnV- centers in nanophotonics]{SnV\textsuperscript{-} centers in nanophotonics}
\begin{figure*}[t]
  \centering
  \includegraphics[width=\textwidth]{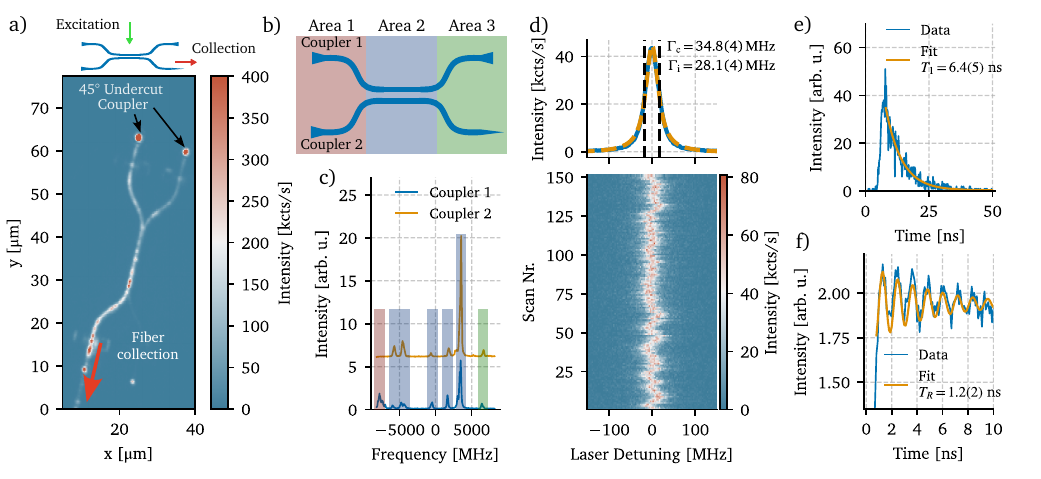}
  \caption{Experimental characterization of SnV$^{-}$ center properties in the fabricated nanophotonic structures.
  \textbf{a)} Confocal image of a directional coupler under off-resonant \SI{520}{\nano\meter} excitation from above, with phonon-sideband (PSB) collection through a tapered-fiber interface.
\textbf{b)} Schematic of a directional coupler divided into three different regions in which emitters can be located.
\textbf{c)} Photoluminescence excitation (PLE) spectra obtained by collecting through the two adjacent TIR couplers.
The approximate spatial region of each SnV$^{-}$ center is inferred from the relative intensities.
\textbf{d)} A cumulative PLE scan with charge-resonant checks between scanned lines yields a spectral-diffusion linewidth of $\Gamma_c = \SI{34.8(4)}{\mega\hertz}$, close to the near-Fourier-transform-limited mean individual linewidth of $\Gamma_i = \SI{28.1(4)}{\mega\hertz}$ obtained from Lorentzian fits to the individual lines.
\textbf{e)} Optical lifetime measurement of an SnV$^{-}$ center in a waveguide structure using a \SI{2}{\nano\second} Gaussian-shaped resonant excitation pulse and PSB detection, yielding $T_1 = \SI{6.4(5)}{\nano\second}$.
\textbf{f)} Optical Rabi oscillations of an SnV$^{-}$ center under a rectangular resonant excitation pulse, showing a Rabi period of \SI{1.2(2)}{\nano\second}.}
\label{fig:experimental-results}
\end{figure*}
To assess the suitability of the freestanding nanophotonic structures as spin--photon interfaces, we investigated whether they provide efficient optical access while preserving the coherence properties of embedded emitters. To characterize the optical properties of SnV$^{-}$ centers in these structures, we fabricated a dedicated sample from electronic-grade diamond. Before nanofabrication, the diamond was implanted with $^{120}\mathrm{Sn}$ ions at an energy of \SI{60}{\kilo\electronvolt} and an ion dose of \SI{2e10}{\text{ions}\per\milli\meter\squared}, yielding an implantation-depth distribution centered at approximately \SI{50}{\nano\meter}. The sample was then annealed for \SI{2}{\hour} at $T\approx\SI{1500}{\celsius}$ to heal implantation-induced lattice damage and promote SnV$^{-}$ formation, which is essential for obtaining favorable emitter properties. This process generated SnV$^{-}$ centers across the sample, with a subset embedded in the fabricated devices.

We first verified the presence of these emitters and assessed their coupling to the tapered-fiber interface by performing a confocal scan under off-resonant excitation at $\lambda_0\approx\SI{520}{\nano\meter}$ through a free-space objective. During the scan, the excitation spot was rastered across the device while the characteristic SnV$^{-}$ phonon-sideband (PSB) fluorescence was collected through the coupled tapered fiber. The PSB is centered around \SI{650}{\nano\meter}; therefore, the fluorescence collected in this measurement probes a wavelength range that is slightly red-shifted relative to the $\approx\SI{620}{\nano\meter}$ ZPL wavelength. As a result, the effective splitting ratio of the directional coupler during PSB collection is expected to differ from that for ZPL emission, as discussed in Supplementary Information \cite{supplementary-information}, Sec.~S3. The resulting image, shown in Fig.~\ref{fig:experimental-results}a, highlights the two upper TIR couplers, where excitation light is efficiently coupled into the waveguide mode. The third coupler is also visible because the coupled-in off-resonant laser excites SnV$^{-}$ centers along the structure, indicating both directional-coupler functionality and emitter coupling to the waveguide mode.

The schematic in Fig.~\ref{fig:experimental-results}b divides the directional coupler into three approximate regions. Emitters in Area 1 couple predominantly to a single output path and are therefore detected mainly through either the TIR coupler labeled 1 or the TIR coupler labeled 2. Emitters in Area 2 lie within the coupling region, so their emission is collected from both output ports but with unequal intensities because the local splitting ratio has not yet reached approximately 50:50. For emitters located towards Area 3, the coupling has evolved further and the collected intensities at the two ports become approximately equal.

This relative intensity distribution provides a qualitative indication of the approximate emitter region within the directional coupler, as inferred from the two-port photoluminescence excitation (PLE) spectrum shown in Fig.~\ref{fig:experimental-results}c. For these measurements, a resonant laser was scanned over the C transition \cite{iwasaki_tin-vacancy_2017} at zero magnetic field. Each peak corresponds to an individual SnV$^{-}$ center within the device, and the colored bars indicate the estimated assignment of the corresponding emitter to Areas 1--3 based on the relative intensities measured at the two collection ports.

We then characterized the optical properties of emitters in photonic structures in more detail. These measurements were performed on a separate but nominally identical implanted and fabricated sample, in which we observed an SnV$^{-}$ center embedded in a waveguide structure. Here, the color center was resonantly excited from top through an objective, while the PSB fluorescence was collected via the tapered-fiber interface. The experimental setup is described in more detail in Supplementary Information \cite{supplementary-information}, Sec.~S2.1. Following the approach in \cite{brevoord_heralded_2024}, we implemented a charge-resonant check (CRC) protocol to account for spectral diffusion and charge-state changes. The threshold for a successful CRC was set to \SI{70}{kcts/s}. The protocol is explained in more detail in Supplementary Information \cite{supplementary-information}, Sec.~S2.2, and the corresponding schematic is shown in Supplementary Information \cite{supplementary-information}, Fig.~3.

The representative PLE spectrum in Fig.~\ref{fig:experimental-results}d exhibits a cumulative full width at half maximum (FWHM) of $\Gamma_c = \SI{34.8(4)}{\mega\hertz}$, while the mean individual linewidth is $\Gamma_i = \SI{28.1(4)}{\mega\hertz}$. The difference between these values can be attributed to spectral diffusion and residual drift of the resonant laser frequency. The measurement was performed using a resonant laser power of \SI{3}{\nano\watt}, measured before the objective, and a CRC was performed between each measured line.

Compared with bulk measurements \cite{bushmakin_two-photon_2024}, the tapered-fiber interface may introduce additional perturbations, including local heating as well as charge- and strain-induced fluctuations, which can increase spectral diffusion. Taken together, these results show that SnV$^{-}$ centers coupled to the waveguide mode retain favorable optical properties, even without post-fabrication annealing, that are comparable to those observed in bulk samples \cite{iwasaki_tin-vacancy_2017}.

To independently verify whether the observed dephasing linewidth approaches the lifetime limit, we further probed the emitter dynamics using a $\Delta t_\mathrm{FWHM}=\SI{2}{\nano\second}$ Gaussian-shaped resonant pulse to measure the optical $T_1$ decay, as shown in Fig.~\ref{fig:experimental-results}e. An exponential fit (see Supplementary Information \cite{supplementary-information}, Sec.~S2.3) yields a lifetime of $T_1=\SI{6.4(5)}{\nano\second}$, corresponding to a lifetime-limited linewidth of $\Gamma_\mathrm{ll}=\SI{25(2)}{\mega\hertz}$. This value is consistent with the measured individual PLE linewidth of $\Gamma_\mathrm{i} = \SI{28.1(4)}{\mega\hertz}$.

Finally, we observed optical Rabi oscillations by driving the SnV$^{-}$ center with a $\Delta t=\SI{20}{\nano\second}$ rectangular resonant pulse at an excitation power of $P_\mathrm{ex}=\SI{6}{\micro\watt}$, measured in front of the objective. The PSB fluorescence emitted during the pulse was collected through the tapered-fiber interface and time-binned, yielding an oscillation period of $T_\mathrm{Rabi}=\SI{1.2(2)}{\nano\second}$ (Fig.~\ref{fig:experimental-results}f), comparable to values reported for bulk samples \cite{iwasaki_tin-vacancy_2017}.

The measurement was fitted using an exponentially damped sine function with a damping time of $\delta = \SI{4.5(9)}{\nano\second}$. The fit function is provided in Supplementary Information \cite{supplementary-information}, Sec.~S2.3. This value is shorter than the previously determined $T_1$, which may be attributed to spectral diffusion during the drive and the resulting detuning between the transition and the laser.

Together, these measurements demonstrate that the applied fabrication process largely preserves the optical properties of embedded SnV$^{-}$ centers, establishing the nanostructures as a suitable platform for SnV$^{-}$-based quantum photonic applications.

  \section{Discussion and outlook}
\label{sec:discussion-outlook}

This work demonstrates that sub-micrometer-spaced freestanding diamond waveguides can be fabricated and released using an entirely dry process while preserving directional-coupler functionality and favorable SnV$^{-}$ optical coherence. The measured devices exhibit nearly balanced power splitting, with a mean coupling ratio of \mbox{$C^\mathrm{meas}=\SI{46\pm16}{\percent}$} at \mbox{$\lambda\approx\SI{637}{\nano\meter}$}, confirming visible-wavelength directional coupling between freestanding triangular diamond waveguides. By avoiding wet processing after release, the dry mask-removal sequence mitigates capillary-force-induced collapse and damage in fragile freestanding nanostructures and should improve process robustness.

The current main limitation is device-to-device variation in evanescent coupling. Improved design, especially of the support structures, together with better control of the angled etch and mask-to-device transfer, will therefore be necessary for reproducible devices in the future, while the longer \SI{738}{\nano\meter} ZPL of SiV$^{-}$ centers could further relax photonic-chip design constraints. Nevertheless, the on-chip beam splitters demonstrated here are expected to provide a key building block for future diamond quantum-photonic circuits.

A direct next verification step is to measure $g^{(2)}(\tau)$ between the two output arms to confirm the routing of photons from a single emitter. Additional on-chip functionality, such as electrodes for Stark or strain tuning of individual defects \cite{bushmakin_two-photon_2024, brevoord_large-range_2025}, would enable spectral alignment of separate SnV$^{-}$ centers and thereby support the routing of indistinguishable photons through on-chip directional couplers. A key next milestone is the observation of Hong--Ou--Mandel interference between such routed photons, ultimately enabling photonic entanglement links between color-center-based quantum nodes.

  \section{Author contributions}
C.S. performed the optical simulations. C.S., O.B., J.E and R.S. contributed to the development of the fabrication procedure. V.B. prepared the SnV$^{-}$-center-enriched bulk diamond samples. C.S. fabricated the nanophotonic devices presented in the manuscript. R.K., T.J., and C.S. performed and analyzed the room-temperature optical measurements. T.J. and C.S. performed and analyzed the low-temperature SnV$^{-}$-center measurements. C.S., T.J., and J.E. wrote the manuscript. V.V. supervised the optical experiments, R.S. supervised the nanofabrication, and J.W. supervised the overall project.

  \section{Acknowledgements}
This research was supported by the German Federal Ministry of Research, Technology and Space (Bundesministerium für Forschung, Technologie und Raumfahrt, BMFTR) through the QR.N initiative (16KIS2207) and the Quanten4KMU (03ZU1110BA) and Quanten4KMU2 (03ZU2110BA) programmes. Additional support was provided through the QuantumBW initiative, jointly funded by the Ministry of Science, Research and Arts Baden-Württemberg (Ministerium für Wissenschaft, Forschung und Kunst Baden-Württemberg, MWK) and the Ministry of Economic Affairs, Labour and Tourism Baden-Württemberg (Ministerium für Wirtschaft, Arbeit und Tourismus Baden-Württemberg), as well as by the MWK through the Center for Integrated Quantum Science and Technology (IQST).
We also thank Victoria Voinkova and Julian Zeller for their valuable contributions to this work through fruitful discussions.

  \printbibliography
\end{refsection}

\begin{refsection}
  \onecolumn
  \setcounter{page}{1}
  \setcounter{section}{0}
  \renewcommand{\thesection}{S\arabic{section}}
  \renewcommand{\theHsection}{supplementary.\arabic{section}}
  \section*{Supplementary Information for \textit{Color-Center-Compatible Freestanding Diamond Directional Couplers for Quantum Photonics}}
  \addcontentsline{toc}{section}{Supplementary Information for Color-Center-Compatible Freestanding Diamond Directional Couplers for Quantum Photonics}

  \section{Directional-coupler fabrication steps}
\label{sup:fabrication}

Fabrication of the directional couplers begins with the definition of a metal hard mask, which is subsequently transferred into bulk single-crystal diamond by multistep RIBE, as detailed in Table~\ref{recipe:bs}. Optical-grade diamond substrates are used for optical performance measurements, whereas electronic-grade diamond substrates are used for SnV$^{-}$-center-related experiments.
Prior to resist coating, the substrates undergo a wet-chemical cleaning procedure to improve resist adhesion and remove contaminants. Each sample is sequentially sonicated in acetone, rinsed in highly purified water, and treated with hot piranha solution ($\ce{H2O2}$:$\ce{H2SO4}$, 1:3 mixture). After a second rinse in highly purified water and nitrogen drying, a soft \ce{O2} plasma etch is applied to activate the surface.
For lithographic patterning, CSAR (AR-P 6200.13, Allresist GmbH) resist is spin-coated onto the cleaned substrate. To provide sufficient conductivity on the insulating substrate, a layer of ESPACER (SHOWA DENKO K.K.) is spin-coated on top. The target structures are then defined by EBL.
The development process begins with a water rinse to dissolve the water-soluble ESPACER layer. The underlying exposed CSAR resist is then developed using the appropriate developer and stopper sequence, followed by an Isopropyl alcohol (IPA) rinse and \ce{N2} drying.
\SI{1}{nm} \ce{Cr} and \SI{80}{nm} \ce{Ti} are deposited on top of the developed resist; after lift-off, these layers form the metal hard-mask pattern.
The thin \ce{Cr} layer is included as a precautionary diffusion barrier between titanium and diamond, reducing the risk of titanium carbide (\ce{TiC}) formation at the interface \cite{tachibana_correlation_1992}. Although \ce{TiC} formation is mainly expected at elevated annealing temperatures, which are not reached during our process, this barrier layer provides an additional safeguard for the hard-mask stack.
The transfer into the bulk material is split into surface-normal and angled etching steps to achieve the targeted directional-coupler profile. First, an initial \ce{Ar} plasma etch is used to refine the metal mask and remove any redeposited material. This is followed by a prolonged anisotropic surface-normal \ce{O2} RIBE step, in which the ion beam is perpendicular to the diamond surface.
Subsequently, the triangular undercuts of the directional couplers are formed by tilting the sample holder to a \SI{45}{\degree} angle of incidence relative to the plasma beam. Continuous in-plane sample rotation is maintained throughout this etching step to promote structural symmetry and uniformity across the substrate.
Finally, the remaining \ce{Ti}/\ce{Cr} hard mask is removed using a highly selective dry etching recipe consisting of a \ce{SF6} plasma step, followed by brief \ce{Ar} physical sputtering and a soft \ce{O2} ICP plasma treatment to yield residue-free directional couplers. The high etch rate of the \ce{SF6} plasma for titanium, together with its low etch rate for diamond, enables selective removal of the titanium layer, whereas the thin chromium barrier layer is removed by brief \ce{Ar} sputtering.

\begin{table}[htbp]
  \centering
  \caption{Fabrication steps for manufacturing directional couplers.}
  \small
  \begin{tabular}{ll}
    \Xhline{2\arrayrulewidth}
    \multicolumn{1}{c}{Process} & \multicolumn{1}{c}{Steps and Parameters}\\ \hline \hline
    Pre-cleaning & \makecell[l]{Sonication in acetone, \SI{4}{min} \\ Rinse in highly purified water\\ Piranha ($\ce{H2O2}$:$\ce{H2SO4}$, 1:3 mixture) at \SI{90}{\celsius}, \SI{5}{min}\\ Rinse in highly purified water\\ Blow-dry with \ce{N2} \\ Soft ICP (PICO plasma cleaner, Diener electronic GmbH \& Co. KG): \SI{50}{\percent} \ce{O2}, \SI{1}{min}} \\ \hline

    Spin coating & \makecell[l]{CSAR \SI{400}{nm} (AR-P 6200.13, Allresist GmbH): \SI{4000}{rpm}, \SI{60}{s}\\ ESPACER: \SI{5000}{rpm}, \SI{60}{s}}\\ \hline

    Lithography & EBL exposure \SI{300}{\micro\coulomb\per\centi\metre\squared}\\ \hline

    Development & \makecell[l]{\SI{30}{s} in highly purified water\\ \SI{1}{min} in CSAR Developer (AR 600-546, Allresist GmbH)\\ \SI{30}{s} in CSAR stopper (AR 600-60, Allresist GmbH)\\ \SI{30}{s} in IPA} \\ \hline

    Mask deposition & \makecell[l]{\SI{1}{nm} \ce{Cr}: \SI{1}{\angstromunit\per\second}\\ \SI{80}{nm} \ce{Ti}: \SI{1}{\angstromunit\per\second}} (UNIVEX 400H, Leybold GmbH) \\ \hline

    Lift-off & \makecell[l]{\SI{2}{min} in CSAR remover (AR 600-71, Allresist GmbH)\\ \SI{2}{s} weak sonication\\ Rinse in IPA \\ Blow-dry with \ce{N2}} \\ \hline

    Surface-normal etch & \makecell[l]{\SI{5}{min} \ce{Ar} metal mask refining etch step, $V_\mathrm{Beam}=\SI{100}{V}$, $I_\mathrm{Beam}=\SI{110}{mA}$\\ \SI{120}{min} surface-normal \ce{O2} etch, $V_\mathrm{Beam}=\SI{50}{V}$, $I_\mathrm{Beam}=\SI{110}{mA}$} (Ionfab, Oxford Instruments plc)\\ \hline

    Angled etch & \makecell[l]{\SI{330}{min} \ce{O2} etching at \SI{45}{\degree} angle of incidence\\ and \SI{2}{rpm} sample in-plane rotation, $V_\mathrm{Beam}=\SI{50}{V}$, \hfill $I_\mathrm{Beam}=\SI{110}{mA}$} (Ionfab, Oxford Instruments plc)\\ \hline

    Mask removal etch & \makecell[l]{\SI{2}{min} \ce{SF6}: $P_\text{HF} = \SI{50}{W}$, $P_\text{ICP} = \SI{300}{W}$\\ \SI{45}{s} \ce{Ar}: $P_\text{HF} = \SI{100}{W}$, $P_\text{ICP} = \SI{300}{W}$\\ \SI{90}{s} \ce{O2} soft ICP: $P_\text{HF} = \SI{0}{W}$, $P_\text{ICP} = \SI{180}{W}$} (PlasmaPro80, Oxford Instruments plc) \\ \hline

    \Xhline{2\arrayrulewidth}
  \end{tabular}
  \label{recipe:bs}
\end{table}

\begin{figure}[H]
  \centering
  \includegraphics[width=\textwidth]{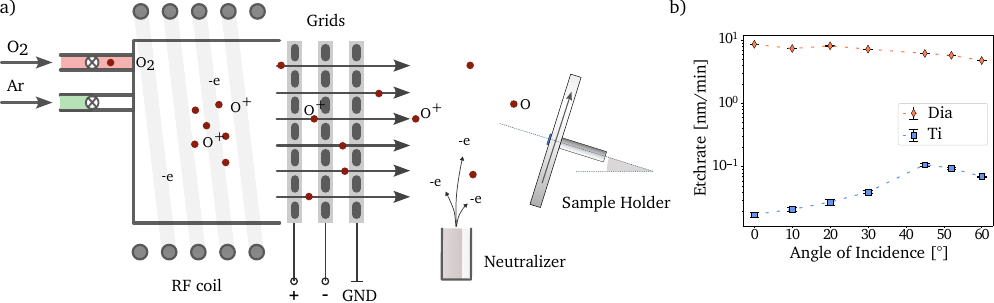}
\caption{\textbf{a)} Schematic of the reactive-ion-beam etching (RIBE) system. The \ce{O2} plasma is generated in the chamber on the right, and ions are extracted and accelerated through grid electrodes to form a directional beam that impinges on the sample. Because the ion source and beamline are mechanically decoupled from the sample stage, the ion energy is set by the beam parameters, while the incidence angle is defined independently by tilting the stage. \textbf{b)} Etch rates of \ce{O2} RIBE in diamond and titanium determined at different angles of incidence under beam parameters of $V_\text{Beam} = \SI{50}{V}$ and $I_\text{Beam} = \SI{110}{mA}$.}
\label{fig:etchrate_angle}
\end{figure}

A crucial component of our fabrication workflow is reactive-ion-beam etching (RIBE), schematically shown in Fig.~\ref{fig:etchrate_angle}a, which enables reproducible angled diamond etching. The system used in this study is the \textit{Ionfab 300+} from \textit{Oxford Instruments}. Because this equipment is not primarily marketed for pure \ce{O2} etching of diamond, process development focused on establishing the etch rates of diamond and masking materials, and therefore their selectivity, over a range of parameters. One key parameter is the beam voltage, which determines the kinetic energy of the ions as they impact the sample.
These measurements showed that the selectivity, defined as the ratio of diamond to titanium etch rates, is significantly enhanced at low energies (\SI{50}{eV}). This behavior is consistent with titanium being etched mainly through physical sputtering, while diamond etching relies chiefly on chemical reactions. Consequently, higher beam voltages promote physical sputtering and reduce selectivity.
The effect of the ion-beam angle of incidence on the etch rates and resulting selectivity was also evaluated.
The resulting data, shown in Fig.~\ref{fig:etchrate_angle}b, indicate that the etch rate of diamond decreases at larger angles measured from the surface normal, while that of titanium increases. This opposing trend is detrimental to the selectivity. Quantitatively, the selectivity decreases by a factor of approximately $10$, dropping from $>500$ at \SI{0}{\degree} to $>50$ at \SI{45}{\degree}.

  \section[SnV- center measurement setup and protocols]{SnV\textsuperscript{-} center measurement setup and protocols}
\subsection{Experimental setup}
\label{sup:experimental_setup}
\begin{figure}[t]
\centering
\includegraphics[width=\columnwidth]{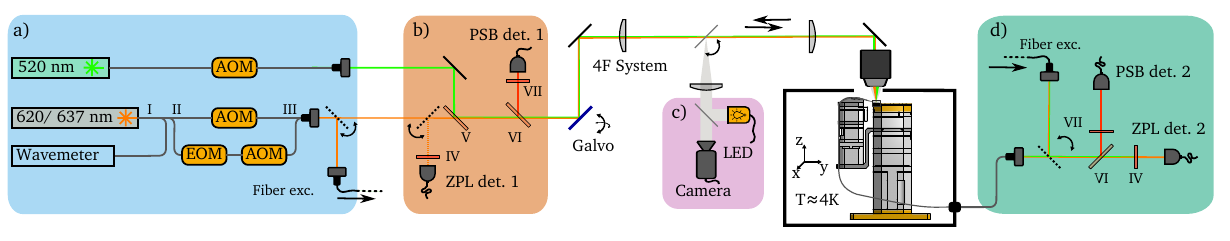}
\caption{Schematic overview of the experimental setup used to characterize diamond photonic chips with embedded SnV$^{-}$ centers. \textbf{a)} Preparation of the off-resonant \SI{520}{\nano\meter} excitation path and the resonant \SI{620}{\nano\meter} control paths using AOMs, an EOM, and fiber-coupled beamsplitters. \textbf{b)} Confocal excitation and collection path, where resonant and off-resonant beams are combined and ZPL/PSB emission can be separated using dichroic mirrors and spectral filters. \textbf{c)} Camera and LED illumination path for top-view imaging of the chip and fiber-coupling alignment. \textbf{d)} Tapered-fiber detection path for collecting fluorescence coupled out of the photonic chip and separating ZPL and PSB emission.}
\label{fig:experimental_setup}
\end{figure}

The cryogenic measurements were conducted in a dilution refrigerator (LPO, QINU GmbH) operated at \SI{4}{\kelvin}. Figure~\ref{fig:experimental_setup} shows the experimental setup, with the three main laser paths illustrated in Fig.~\ref{fig:experimental_setup}a. The green \SI{520}{\nano\meter} laser is pulsed, and its power is controlled using a fiber-coupled acousto-optic modulator (AOM). The \SI{620}{\nano\meter} resonant laser (DLC TA-SHG PRO, Toptica Photonics AG) is first split by a 99/1 fiber-coupled beamsplitter (I), with the 1\% output sent to a wavemeter for wavelength monitoring. The remaining 99\% output is sent to an additional fiber-coupled 75/25 beamsplitter (II), which separates the resonant light into two control paths. The lower-power output is guided to a fiber-coupled AOM for power tuning and modulation. All AOMs are driven by external radio-frequency (RF) drivers controlled by a field-programmable gate array (FPGA; OPX+, QM Technologies Inc.) with a clock-cycle resolution of \SI{4}{\nano\second}.

The higher-power output is sent through an electro-optic modulator (EOM) to generate high-resolution arbitrary waveforms and then through an AOM to suppress unwanted transmission from the EOM pulse and tune the laser power. The EOM is locked to zero transmission using a lock-in amplifier in combination with a PID controller. The EOM is driven by an arbitrary waveform generator (AWG; PPG 512, PicoQuant GmbH), which generates triggered arbitrary waveforms with a temporal resolution of \SI{200}{\pico\second}.

Before recombination, the polarizations of the two resonant paths are matched using quarter- and half-wave plates. The two resonant paths are then combined using a 75/25 fiber-coupled combiner (III). A flip mirror is used to switch between excitation through the tapered fiber and confocal excitation from above through an objective.

The resonant and off-resonant laser paths are then combined using a \SI{550}{\nano\meter} long-pass dichroic mirror (V), as shown in Fig.~\ref{fig:experimental_setup}b). For collection of resonant zero-phonon-line (ZPL) emission from above, a flip mirror and a \SI{620}{\nano\meter} narrow-band filter (IV) can be inserted while the resonant excitation path is blocked. A sharp \SI{625}{\nano\meter} short-pass dichroic mirror (VI) separates ZPL emission from phonon-sideband (PSB) emission. An additional \SI{635}{\nano\meter} long-pass filter (VII) suppresses residual laser light.

The combined optical paths are guided to a dual-axis galvo mirror integrated into a 4f system, enabling confocal raster scanning of the sample through the objective inside the cryostat.
A flippable pellicle beamsplitter connects the confocal microscope to a camera and LED illumination path, as shown in Fig.~\ref{fig:experimental_setup}c, allowing top-view imaging of the chip and facilitating fiber coupling.
The sample is positioned at the focus of the objective, which is mounted on the \SI{4}{\kelvin} stage, using a stack of piezo steppers that provide alignment in the $x$, $y$, and $z$ directions.
A Z-shaped bracket with three smaller piezo steppers is used to align the tapered fiber relative to the chip while also allowing the combined tapered-fiber--chip interface to be moved as a single unit relative to the objective.

Fluorescence coupled into the tapered fiber is guided out of the cryostat through the optical fiber and directed to a separate detection setup, as shown in Fig.~\ref{fig:experimental_setup}d.
As in the confocal detection path, a sharp \SI{625}{\nano\meter} short-pass dichroic mirror (VI) separates the ZPL and PSB emission.
For ZPL detection, an additional \SI{620}{\nano\meter} narrow-band filter (IV) suppresses residual off-resonant light, while a \SI{635}{\nano\meter} long-pass filter is used for PSB detection.
\subsection{Charge-Resonant-Check Protocol}
\label{sup:crc}
\begin{figure}[h]
\centering
\includegraphics[width=\columnwidth]{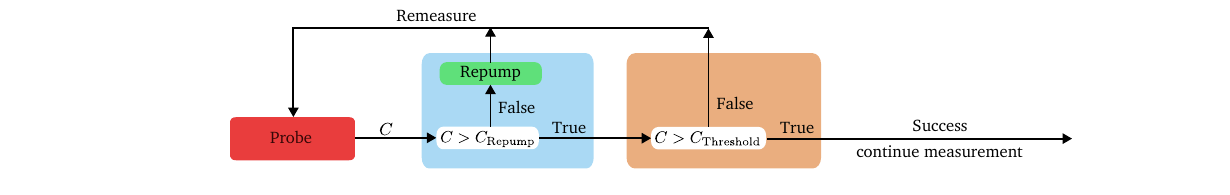}
\caption{Charge-resonant-check (CRC) sequence used to verify the SnV$^{-}$ charge state before each measurement. A resonant probe measures counts $C$. If $C>C_{\mathrm{Repump}}$, the experiment proceeds; otherwise, an off-resonant repump pulse is applied and the probe is repeated. The measurement starts once $C>C_{\mathrm{Threshold}}$.}
\label{fig:crc}
\end{figure}

The charge-resonant check (CRC) \cite{brevoord_heralded_2024}, illustrated in Fig.~\ref{fig:crc}, is used to verify the SnV$^{-}$ charge state before each measurement sequence.
First, a short resonant probe pulse is applied, and the detected photon counts $C$ are compared with the repump threshold $C_{\mathrm{Repump}}$.
If $C>C_{\mathrm{Repump}}$, the emitter is considered to be in the desired charge state and the sequence proceeds directly to the measurement.
If the count level is too low, an off-resonant repump pulse is applied to recover the charge state, followed by a second resonant probe.
The measurement is started only when the subsequent count rate exceeds the threshold $C_{\mathrm{Threshold}}$; otherwise, the CRC cycle is repeated until the condition is met.
In our experiments, the repump threshold was set to $C_{\mathrm{Repump}}=\SI{10}{kcts/s}$, and the measurement threshold was set to $C_{\mathrm{Threshold}}=\SI{70}{kcts/s}$.
\subsection{Analysis and Fit Models}
\label{sup:fit}
The optical lifetime $T_1$ is extracted by fitting the time-resolved fluorescence decay with a single-exponential decay model,
\begin{equation*}
I(t) = A \exp\left(-\frac{t}{T_1}\right) + B.
\end{equation*}
Here, $I(t)$ is the detected fluorescence intensity as a function of time, $A$ is the decay amplitude, $B$ accounts for a constant background offset, and $T_1$ is the excited-state lifetime. The fit yields $T_1=\SI{6.4(5)}{\nano\second}$ for the SnV$^{-}$ center discussed in the main text.

The optical Rabi oscillations are analyzed using an exponentially damped sinusoidal model,
\begin{equation*}
I(t) = A \sin\left(\frac{2\pi t}{T_\mathrm{Rabi}} + B\right) \exp\left(-\frac{t}{\delta}\right) + C.
\end{equation*}
Here, $I(t)$ is the detected fluorescence intensity, $A$ is the oscillation amplitude, $T_\mathrm{Rabi}$ is the Rabi period, $B$ is the phase offset, $\delta$ is the damping time of the oscillations, and $C$ accounts for a constant background offset.
The fit yields a Rabi period of $T_\mathrm{Rabi}=\SI{1.16(1)}{\nano\second}$ and a damping time of $\delta=\SI{4.5(9)}{\nano\second}$.
The damping time is shorter than the independently measured lifetime $T_1=\SI{6.4(3)}{\nano\second}$, which may originate from residual detuning between the laser and the optical transition during the drive, for example due to spectral diffusion.

  \section{Photonic measurements and simulations}
\label{sup:photonic_measurements}
\subsection*{TIR coupler}
The TIR couplers provide free-space optical access to the in-plane diamond waveguides. They are fabricated together with the suspended photonic circuit during the angled RIBE step and form an inclined diamond interface at the end of the waveguide. Light focused from above is redirected by this interface into the plane of the chip and subsequently coupled into the guided mode through the tapered waveguide section. Conversely, guided light can be redirected or scattered out of plane and collected with the same objective.

\subsection*{Comparing measured geometry with simulations}
To better understand the coupling measured in the fabricated devices, we perform eigenmode simulations using the device geometry extracted from SEM measurements. The simulated structures are constructed from the measured waveguide widths, \mbox{$w_\mathrm{wg}^\mathrm{meas}$}, and waveguide separations, \mbox{$d_\mathrm{c}^\mathrm{meas}$}, summarized together with the design parameters in Table~\ref{tab:sem_measured_couplers}. For the outer sidewalls of the directional coupler, we assume an undercut angle \mbox{$\alpha_\mathrm{uc}^\mathrm{meas}=\SI{42\pm 2}{\degree}$}, equal to the reference angle measured on a waveguide fabricated with the same process. The inner undercut angle $\alpha_\mathrm{in}$ denotes the sidewall angle on the two waveguide sides that face each other across the coupling gap. This angle was not measured directly and may be steeper than the outer angle because these opposing sidewalls are affected by shadowing during the angled etch.

\begin{table}[ht]
  \centering
  \caption{Measured directional-coupler geometry and mean coupling for each measured coupler with different numbers of supports in the coupling region, $N$. The coupling uncertainty is the standard deviation over the individual data points.}
  \label{tab:sem_measured_couplers}
  \begin{tabular}{ccccccccc}
    \hline
    $d_\mathrm{c}^\mathrm{des}$ [nm] & $d_\mathrm{c}^\mathrm{meas}$ [nm] & $w_\mathrm{wg}^\mathrm{meas}$ [nm] & $C_\mathrm{mean}^\mathrm{meas}$ [\%] & $C_{N=1}^\mathrm{meas}$ [\%] & $C_{N=2}^\mathrm{meas}$ [\%] & $C_{N=3}^\mathrm{meas}$ [\%] & $C_{N=4}^\mathrm{meas}$ [\%] & $C_{N=5}^\mathrm{meas}$ [\%] \\
    \hline
    \num{200} & \num{221 \pm 9} & \num{287 \pm 9} & \num{46.36 \pm 16.29} & 43.75 & 72.34 & 27.63 & 41.03 & 47.06 \\
    \num{250} & \num{279 \pm 9} & \num{283 \pm 9} & \num{6.26 \pm 4.70} & 1.84 & 2.83 & 4.04 & 12.20 & 10.38 \\
    \num{300} & \num{335 \pm 9} & \num{294 \pm 9} & \num{7.71 \pm 1.83} & 7.50 & 9.91 & 6.70 & 9.09 & 5.34 \\
    \hline
  \end{tabular}
\end{table}

\begin{figure}[ht]
  \centering
  \includegraphics[width=\columnwidth]{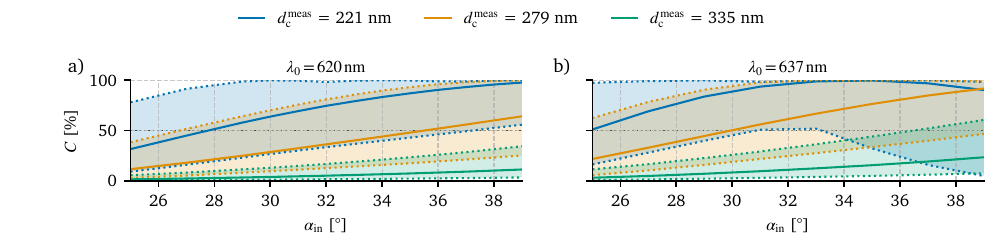}
\caption{Eigenmode simulations of the coupling ratio based on device parameters extracted from SEM measurements at \textbf{a)} $\lambda_0\approx\SI{620}{nm}$ and \textbf{b)} $\lambda_0\approx\SI{637}{nm}$. The coupling is calculated for different assumed inner undercut angles in the coupling region, while the outer undercut angle, waveguide widths, and waveguide separation are taken from the SEM measurements. The colored bands indicate the resulting coupling uncertainty under the worst-case combination of all assumed geometric errors, highlighting the sensitivity of the device response to fabrication variations.}
\label{fig:sim_measurement_results}
\end{figure}

We therefore sweep the inner undercut angle in the simulations to quantify its influence on the coupling. In addition, we simulate the worst-case combinations of the geometric deviations within the SEM measurement uncertainties to estimate the resulting uncertainty in the simulated coupling. Fig.~\ref{fig:sim_measurement_results} shows the results for the three measured waveguide separations $d_\mathrm{c}^\mathrm{meas}$ as a function of the assumed inner undercut angle; the colored bands indicate the spread caused by the geometric uncertainty. The simulations show that nanometer-scale variations in waveguide geometry and degree-scale variations in sidewall angle can strongly affect the coupling. Consequently, a one-to-one assignment of the measured coupling values to a unique simulated geometry is challenging. Nevertheless, for specific inner undercut angles, the measured coupling values fall within the simulated uncertainty bands.

\subsection*{Loss analysis of the directional couplers}

\begin{figure}[ht]
\centering
\includegraphics[width=\columnwidth]{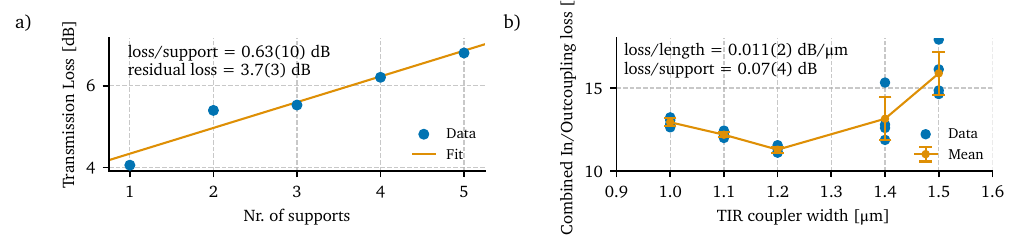}
\caption{\textbf{a)} Optical loss contribution as a function of the number of supports connecting the coupling region for the five measured directional couplers, after subtracting the expected loss from the TIR couplers. \textbf{b)} Extracted optical loss contribution of the TIR couplers for different TIR-coupler widths. Losses from waveguide propagation and support structures were obtained by fitting measurements from waveguides with different lengths and numbers of supports and were subtracted from the displayed coupler loss. The lowest loss is observed for a TIR-coupler width of \SI{1.2}{\micro\meter}, corresponding to an estimated single-coupler loss of $L_\mathrm{TIRC}=\SI{5.5}{\decibel}$.}
\label{fig:loss_couplers}
\end{figure}

For the photonic transmission measurements of the directional couplers, we estimate the loss introduced by the coupler geometry and by the mechanical supports. We describe all contributions as optical losses in decibels, defined relative to the input power as $-10\log_{10}(P_\mathrm{out}/P_\mathrm{in})$. With this convention, positive values correspond to attenuation. The loss through a directional coupler is modeled as
\begin{equation}
L_{\mathrm{dc}} = L_\mathrm{TIRC} + n_{\mathrm{supports}} L_{\mathrm{support}} + L_{\mathrm{residual}},
\label{eq:directional_coupler_loss_model}
\end{equation}
where $L_\mathrm{TIRC}$ is the combined in- and out-coupling loss of the TIR couplers, $L_{\mathrm{support}}$ is the loss per support within the coupling region, and $L_{\mathrm{residual}}$ accounts for remaining loss channels, such as scattering from supports outside the coupling region and propagation loss caused by waveguide surface roughness. The five measured directional couplers with approximately 50:50 power splitting contain between one and five supports in the coupling region. Fitting Eq.~\ref{eq:directional_coupler_loss_model} to these data therefore yields the support-dependent loss shown in Fig.~\ref{fig:loss_couplers}a. However, this measurement alone does not separate residual propagation and scattering losses from the in- and out-coupling losses.

To independently characterize the TIR couplers, we fabricated calibration waveguides with different lengths, $l_\mathrm{wg} \in \{\SI{50}{\micro\meter}, \SI{75}{\micro\meter}, \SI{100}{\micro\meter}\}$, and different numbers of supports, $n_\mathrm{supports} \in \{2,3,4,5\}$. Their total loss is described by
\begin{equation}
L_{\mathrm{wg}} = L_\mathrm{TIRC} + n_{\mathrm{supports}} L_{\mathrm{support,wg}} + l_{\mathrm{wg}} L_{\mathrm{per-\mu m,wg}},
\label{eq:waveguide_loss_model}
\end{equation}
which allows us to extract the loss per support, $L_{\mathrm{support,wg}}$, and the propagation loss per micrometer, $L_{\mathrm{per-\mu m,wg}}$. We then use a second set of calibration waveguides, in which the TIR-coupler width is varied, to determine the combined in- and out-coupling loss after subtracting the independently extracted support and propagation losses. The resulting TIR-coupler-width dependence is shown in Fig.~\ref{fig:loss_couplers}b. The lowest loss is obtained for a TIR-coupler width of \SI{1.2}{\micro\meter}. Because input coupling from the focused Gaussian laser beam and output collection from the guided mode are not necessarily reciprocal in the experiment, the extracted value represents the combined in- and out-coupling loss. Assuming equal in- and out-coupling efficiencies only as a reference, the best measured TIR-coupler width corresponds to an estimated single-TIR-coupler loss of $L_\mathrm{TIRC} = \SI{5.5}{\decibel}$, equivalent to a transmission of $T_\mathrm{TIRC} = \SI{28}{\percent}$. The directional couplers use \SI{1}{\micro\meter}-wide TIR couplers; for this width, the combined in- and out-coupling loss is $L_\mathrm{TIRC} = \SI{12.9\pm0.3}{\decibel}$. Applying the corresponding coupling correction yields a residual directional-coupler loss of $L_{\mathrm{residual}} = \SI{3.7\pm0.3}{\decibel}$ and a support loss of $L_{\mathrm{support}} = \SI{0.6\pm0.1}{\decibel}$ per support.

\subsection*{FDTD simulations}
The FDTD simulations were performed using Lumerical FDTD to provide a three-dimensional verification of the directional-coupler design at a wavelength of $\lambda_0=\SI{620}{\nano\meter}$. The lithography mask layout was imported into the simulation and extruded using the assumed undercut angle to obtain the full device geometry. Diamond was modeled with a nondispersive refractive index of $n=2.41$, and all simulation boundaries were set to perfectly matched layers (PMLs). In the coupling region, we used a mesh-override region with grid spacings of \SI{15}{\nano\meter} along $x$, the waveguide propagation direction, and \SI{5}{\nano\meter} along both $y$ and $z$. The mesh-override region included at least \SI{200}{\nano\meter} of padding around the structure in the $y$ and $z$ directions to accurately resolve the triangular waveguide cross section and suppress staircasing artifacts. Light was launched into the input port $p_0$ with a mode source exciting the fundamental TE-like waveguide mode. At the two output ports, $p_1$ and $p_2$, mode-expansion monitors were used to extract the powers $P_1$ and $P_2$ carried by the corresponding fundamental TE-like modes. The coupling ratio was then calculated as $C=P_2/\left(P_1+P_2\right)$. A two-dimensional electric-field monitor in the device plane was used to visualize the field distribution shown in Fig.~2b of the main text. To check numerical convergence, the simulation was repeated for different mesh resolutions, which yielded comparable coupling ratios. The FDTD simulation gives a slightly overcoupled response of $C^\mathrm{sim}=\SI{51}{\percent}$. This small deviation from the ideal 50:50 value can be attributed to residual coupling in the transition regions before and after the nominal coupling section, as well as to numerical differences between the full three-dimensional FDTD model and the eigenmode-based design calculation.

  \printbibliography[title={Supplementary References}]
\end{refsection}

\end{document}